\documentclass[aip,reprint,amsmath,amssymb]{revtex4-2}

\usepackage{booktabs}
\usepackage{siunitx}
\usepackage{graphicx}
\usepackage{dcolumn}
\usepackage{bm}
\usepackage[utf8]{inputenc}
\usepackage[T1]{fontenc}
\usepackage{mathptmx}
\usepackage{etoolbox}
\usepackage{hyperref}
\usepackage{svg}
\usepackage{multirow}
\usepackage{array}
\usepackage{tabularx}
\usepackage{color}

\definecolor{mygreen}{rgb}{0,0.5,0}
\definecolor{mygrey}{rgb}{0.5,0.5,0.5}
\definecolor{myred}{rgb}{0.75,0,0}
\definecolor{myblue}{rgb}{0,0,0.75}
\definecolor{mymagenta}{cmyk}{0,1,0,0.12}
\definecolor{mycyan}{cmyk}{1,0,0,0.12}
\definecolor{myorange}{rgb}{1.,0.5,0}
\definecolor{myviolet}{rgb}{0.6,0.15,0.6}
\definecolor{mybrown}{cmyk}{0,0.50,1,0.41}

\DeclareSIUnit\torr{Torr}
\DeclareSIUnit\amagat{amg}
\makeatletter
\def\@email#1#2{%
 \endgroup
 \patchcmd{\titleblock@produce}
  {\frontmatter@RRAPformat}
  {\frontmatter@RRAPformat{\produce@RRAP{*#1\href{mailto:#2}{#2}}}\frontmatter@RRAPformat}
  {}{}
}%
\makeatother

\newcommand{\SSSIHL}{FEMTO FabULLAS, CRIF, Department of Physics, Sri Sathya Sai Institute of Higher Learning, Prasanthi Nilayam, Andhra Pradesh, 515134, India}

\begin{document}

\title{Fabrication and study of femtosecond laser micromachined few-mode elliptical core waveguides}

\author{Prajal Chettri}
\affiliation{\SSSIHL}

\author{Shailesh Srivastava}
\affiliation{\SSSIHL}

\email{prajalchettri@sssihl.edu.in}

\date{\today}

\begin{abstract}
Femtosecond laser micromachining (FLM) fabricated waveguides inherently form elliptical cores due to differences in focal spot size and the Rayleigh range of the microscope objective. Consequently, it is essential to study their propagation characteristics, which differ from those of conventional circular-core waveguides. In this work, we present the results of a parametric optimization of these waveguides to identify fabrication parameters that lead to minimal loss. A propagation loss characterization study revealed that, for a laser wavelength of 1030~nm, a pulse width of $\sim$300~fs, a pulse energy of 600~nJ, a scan speed of 2~mm/s, and a repetition rate of 100~kHz, a transparent and micro-bubble-free waveguide with a propagation loss of $\sim$0.4~dB/cm was formed. The modal analysis further demonstrated that the V-number depends on the core aspect ratio. The waveguide modes were compared with computationally generated modes, revealing a correlation that aligns well with existing literature. 
\end{abstract}


\maketitle
\section{\label{sec:intro}Introduction}
Femtosecond laser micromachining (FLM) is a versatile tool that has been used extensively for the fabrication of various integrated photonic devices. The high precision of femtosecond pulses ensures minimal heat diffusion, allowing controlled modification of the material with sub-micrometer accuracy. By focusing the laser beam inside the material, continuous waveguides can be formed. One of the key advantages of FLM is its ability to create three-dimensional structures, as the laser can be focused at any point inside the bulk of the material. This, coupled with minimal thermal effects, makes FLM ideal for fabricating complex integrated photonic circuits. Applications of FLM-fabricated waveguides are vast, ranging from photonic circuits and sensors to quantum optics and telecommunications.

Optical waveguides fabricated in the low pulse energy regime in fused silica increase the refractive index in the laser-irradiated region~\cite{ref1,ref2}. This forms the basis for the waveguiding mechanism, with a higher-index core and an unirradiated lower-index cladding. Due to the inherent third-decimal change in refractive index, the waveguides formed are very similar to standard optical fibers, as they are weakly guiding ($n_{\mathrm{core}} \sim n_{\mathrm{clad}}$). It has been shown~\cite{ref3} that using high repetition rates results in heating effects that make the core circular. Without these heating effects, the core would be elliptical due to top-surface-induced spherical aberrations as well as the difference in the Rayleigh range (which determines the vertically affected zone) compared to the spot diameter (which determines the horizontally affected zone). \par

The molecular structure of fused silica can be described as a continuous yet random network of cross-linked SiO$_2$ in a tetrahedral arrangement, with a network of closed paths of Si--O bonds forming ring structures. As confirmed by Pasquarello \textit{et al.}~\cite{ref4}, the ring sizes in fused silica range from three- up to nine-membered rings, with five- and six-membered structures constituting the majority. The two Raman peaks at 495 and 605~cm$^{-1}$ correspond to the Raman-active symmetric breathing modes of the oxygen atoms in the four- and three-membered ring structures, respectively, as reported in the studies of Galeener \emph{et al.}, which were later modeled and confirmed by Pasquarello \emph{et al.} and Car \emph{et al.}~\cite{ref5,ref6,ref7,ref8}. The broad peaks in the Raman spectrum of fused silica arise from vibrations of the cross-linked glass network, and their broadness indicates a wide distribution of Si--O--Si bond angles. The Raman spectra of femtosecond-laser-modified regions in fused silica show a large increase in the peaks at 490 and 605~cm$^{-1}$, attributed to four- and three-membered ring structures in the silica network, indicating that densification occurs after exposure to femtosecond laser pulses. These results are consistent with the formation of a localized plasma induced by the laser pulse~\cite{ref9}.\par

Recently, Ryo Imai \textit{et al.}~\cite{ref10} demonstrated the effective use of a 1030~nm laser in fused silica and determined the parameters required for obtaining low-loss waveguides. Their study utilized a slit beam-shaping technique~\cite{ref11} to achieve circular cross-sections for single-mode waveguides, all written at a repetition rate of 100~kHz. They showed that pulse widths needed to be less than 400~fs and pulse energies $\leq 0.8~\mu$J for low-loss waveguide fabrication. An advantage of working at 1030~nm is that higher pulse energies are typically available at the fundamental wavelength compared to the second-harmonic generation (SHG) output in these laser sources, making the process more cost-effective and easier to implement.\par

It is interesting to note that most reports in the literature correct the ellipticity of the waveguide cross-section using various strategies. For example, Imai \textit{et al.}~\cite{ref10} utilized an optical slit to modify the spatial profile of the laser beam. Lapointe \textit{et al.}~\cite{ref12} employed a cylindrical lens telescope to produce an astigmatic beam for beam shaping, while Cruz \textit{et al.}~\cite{ref13} demonstrated the use of astigmatic elliptical beams generated using spatial light modulators (SLMs). These approaches are typically adopted to minimize birefringence, ensure uniform mode confinement, and increase coupling efficiency with other optical components. However, there are no reports on detailed studies of the modal and loss characteristics of femtosecond laser micromachined elliptical-core waveguides and their application in biosensing.\par

In this work, we report the fabrication and optimization of optical waveguides in fused silica using femtosecond laser micromachining. The ellipticity of the waveguide cross-sections and their widths are studied as a function of various fabrication parameters. Furthermore, the propagation loss of these waveguides is characterized for guiding visible wavelengths using the cut-back method. Finally, the output modes of these few-mode elliptical-core waveguides are studied and correlated with theoretical models reported in the literature.

\section{Methods}

\subsection{Fabrication of Waveguides}

With the aim of writing waveguides fine-tuned to attain optimal guiding characteristics at a wavelength of 532~nm, femtosecond laser pulses from a ytterbium-doped fiber laser with a central wavelength of 1030~nm were focused using a microscope objective with a large working distance and a numerical aperture (NA) of 0.42. The laser was operated at a repetition rate of 100~kHz for waveguide fabrication, as this regime has been shown to exhibit lower propagation losses~\cite{ref10}. For local optimization, pulse energies ranging from 300~nJ to 1000~nJ in steps of 100~nJ and translation speeds of 0.5~mm/s, 1~mm/s, and 2~mm/s were selected.

Waveguides were written at a depth of 400~$\mu$m using parallel polarization (parallel to the writing direction) in a fused silica chip of thickness 1~mm and length 2~cm (FOCtek Photonic Inc., China). This depth is well suited for coupling into and out of the microchannels used in this work. After waveguide fabrication, the end facets were cut using a diamond glass cutter (Baincut LSS, Chennai Metco) and subsequently polished using a glass grinder and polishing machine (Bainpol VTD, Chennai Metco).

\subsection{Characterization of the Waveguides}

Bright-field microscopy was employed as a qualitative tool to assess the laser-induced refractive index modification and the uniformity of the fabricated waveguides. A transmission optical microscope (Laben Instruments, BM--3000~FLT) was used for this purpose. The waveguide cross-sections and widths were imaged at magnifications of 100$\times$ and 400$\times$, respectively. For calibration, a USAF resolution test target was imaged under identical magnification conditions. Using ImageJ software, the semi-major and semi-minor axes of the waveguide cross-sections, as well as the widths of the waveguides, were determined. \par

\begin{figure}[t]
\centering
\includegraphics[width=\columnwidth]{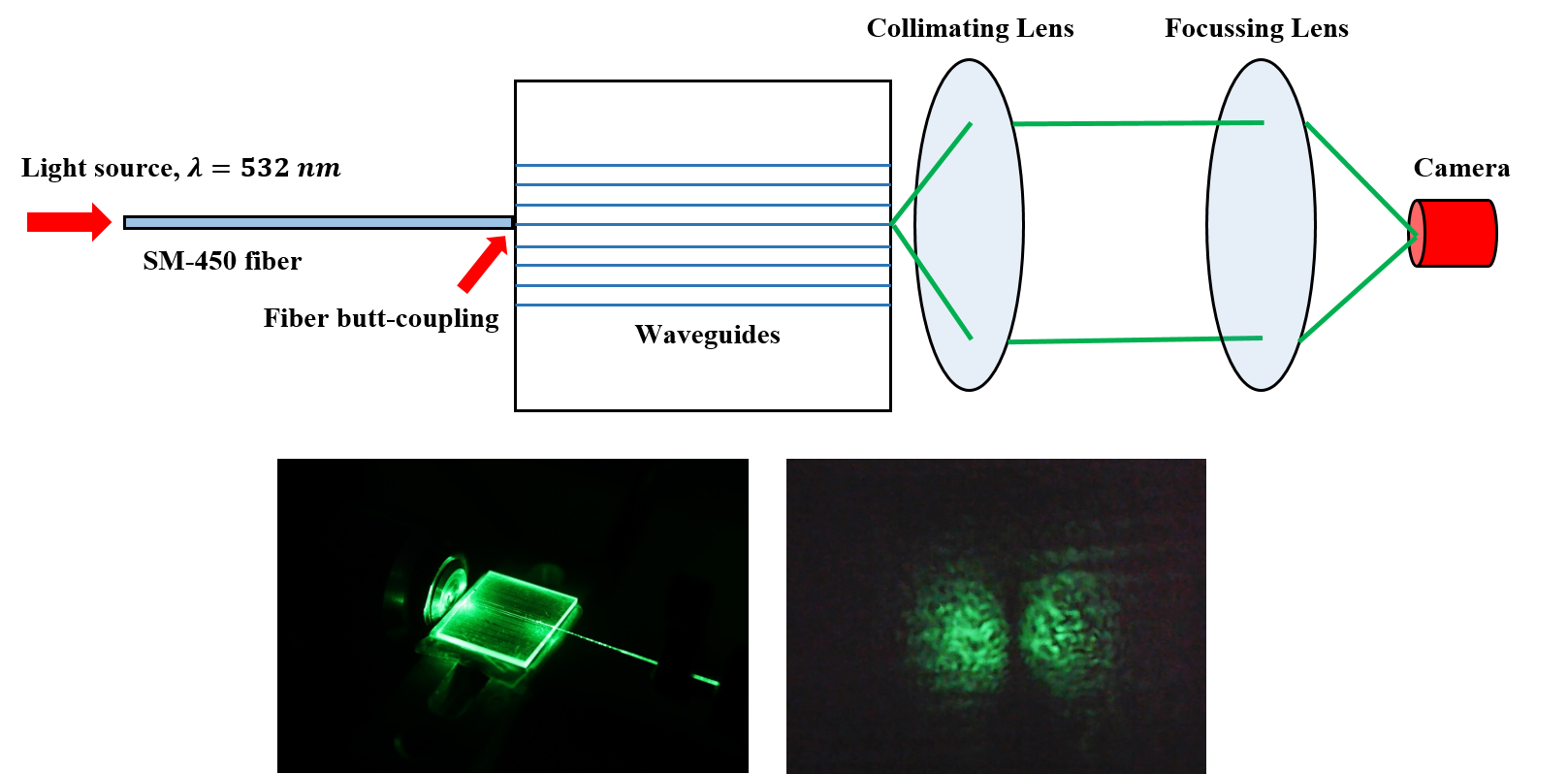}
\caption{Experimental schematic for waveguide characterization}
\label{fig:Fig1}
\end{figure}

As shown in Fig.~\ref{fig:Fig1}, the waveguide characterization setup consisted of a diode laser ($\lambda = 532$~nm), two microscope objectives of identical numerical aperture (0.25) and magnification (10$\times$), a standard single-mode fiber optimized for 532~nm (SM-450) with both ends cut, cleaned, and cleaved, a focusing lens ($f = 10$~cm), and a CMOS camera (Logitech C-170). Both microscope objectives were mounted on five-axis positioners (translation along $X$, $Y$, and $Z$ axes, and tilt in the $YZ$ and $XZ$ planes) to enable precise alignment and positioning. Each end of the fiber was held in a fiber chuck and mounted on separate five-axis positioners.\par

The output of the diode laser was focused into the fiber by adjusting the five-axis positioners of the input objective and the fiber. The transmitted power through the fiber was maximized by fine-tuning the positioners at the input end of the fiber and the first microscope objective, and the output power was measured using a power detector (Ophir Nova~II). Once the maximum output power was achieved, a platform mounted on a three-axis translation stage was placed between the output fiber tip and the collimating lens (objective~2). The femtosecond laser micromachined waveguides fabricated in the fused silica chip were then positioned on this platform.\par

The propagation loss was evaluated by measuring the difference in output power from two identical waveguides with lengths of 17~mm and 9~mm. The loss was calculated using the following expression:
\begin{equation}
\mathrm{Loss}\;(\mathrm{dB/cm}) = \frac{10}{L_1 - L_2}\,\log\!\left(\frac{P_{L_2}}{P_{L_1}}\right),
\end{equation}
where $L_1$ and $L_2$ are the lengths of the two waveguides ($L_1 > L_2$), and $P_{L_1}$ and $P_{L_2}$ are the corresponding output powers measured from waveguides of lengths $L_1$ and $L_2$, respectively.

\section{Results and discussion}
\subsection{Cross-sectional ellipticity dependence on laser fabrication parameters}
\begin{figure}[t]
\centering
\includegraphics[width=\columnwidth]{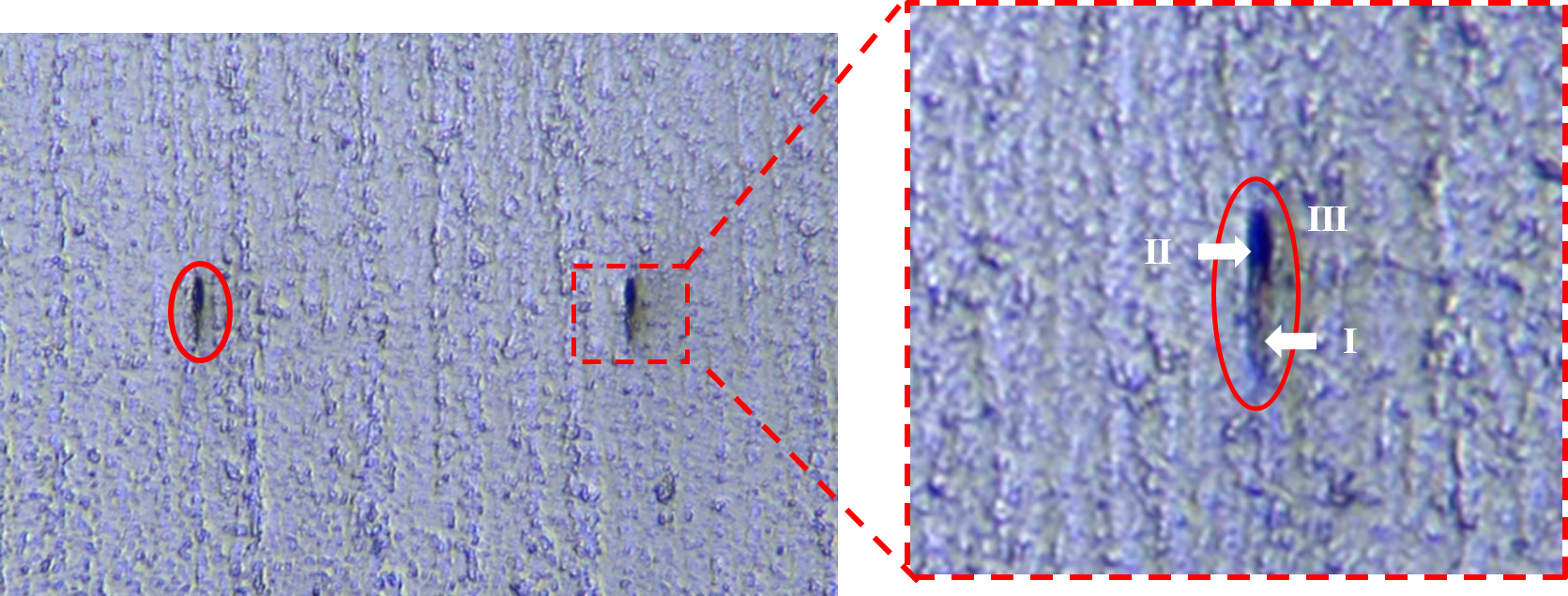}
\caption{Optical microscope image of the waveguides depicting the dark and bright zones within the cross-section.}
\label{fig:Fig2}
\end{figure}

As discussed earlier, femtosecond laser micromachined optical waveguides inherently form elliptical cross-sections. Upon close observation, the elongated structures exhibit distinct bright and dark regions in the waveguide cross-sections, as shown in Fig.~\ref{fig:Fig2}. According to Tan \textit{et al.}~\cite{ref14}, the dark region corresponds to a negative refractive index change and therefore does not support light guiding. In contrast, the bright region corresponds to a positive refractive index change and acts as the effective waveguide core~\cite{ref14,ref15,ref16,ref17,ref18,ref19}. \par
The formation of these distinct refractive index regions is attributed to the redistribution of laser energy along the beam propagation direction. Due to the high intensities involved, plasma generation can occur in the irradiated region before the geometrical focus~\cite{ref20,ref21}. When the temperature in this pre-focal region reaches the softening point of silica, thermal expansion is triggered, leading to rarefaction. This results in a negative density change and, consequently, a decrease in the refractive index. Conversely, the most widely accepted explanation for a positive refractive index change is densification caused by fs-laser-induced reorganization of the silica structural network, characterized by an increase in three- and four-membered ring structures and a reduction in the average Si--O--Si bond angle. \par

\begin{figure*}[t]
\centering
\includegraphics[width=\textwidth]{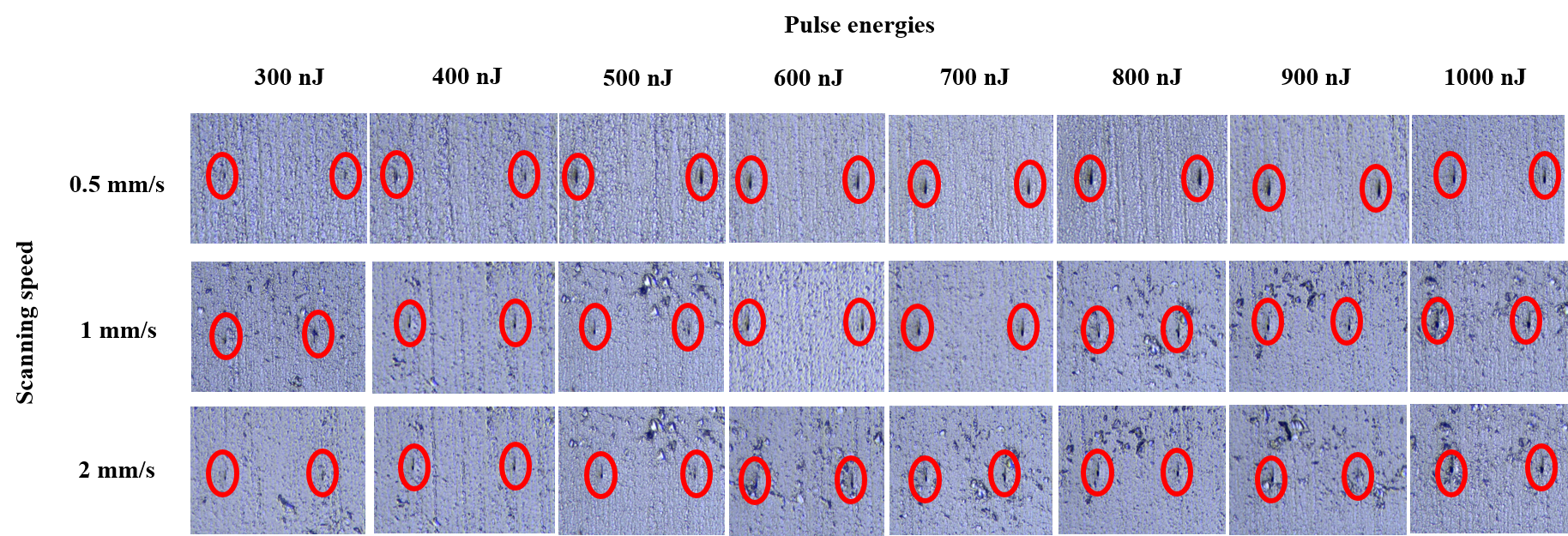}
\caption{Cross-sectional optical microscope images of optical waveguides fabricated with different pulse energies and scanning speeds.}
\label{fig:Fig3}
\end{figure*}

Bright-field microscope images of the cross-sections of femtosecond laser micromachined elliptical-core waveguides are shown in Fig.~\ref{fig:Fig3}. To quantify the degree of ellipticity in the cross-section, the eccentricity $e$ is defined as the ratio of the distance from the center of the ellipse to either focus to the length of the semi-major axis. Mathematically, the eccentricity is expressed as
\begin{equation}
e = \frac{\sqrt{a^2 - b^2}}{a}, \qquad 0 \leq e < 1,
\end{equation}
where $a$ and $b$ denote the semi-major and semi-minor axes of the elliptical waveguide core, respectively. \par
\begin{figure}[t]
\centering
\includegraphics[width=\columnwidth]{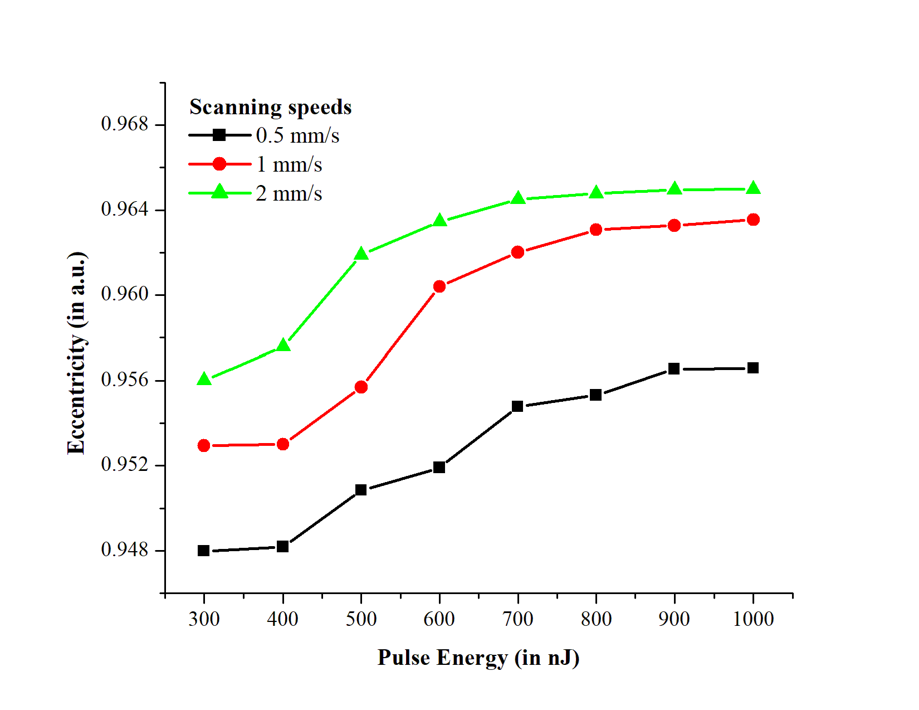}
\caption{Eccentricities as a function of pulse energies for different scanning speeds.}
\label{fig:Fig4}
\end{figure}

The influence of pulse energy and the number of irradiated pulses on the eccentricity was systematically investigated. Only a marginal increase in eccentricity was observed with increasing pulse energy, which can be attributed to the absence of significant heating effects at lower repetition rates. Furthermore, lower eccentricities were observed at reduced scan speeds, as the larger number of irradiated pulses at a given point (higher laser fluence) resulted in a more substantial refractive index modification in both transverse dimensions, thereby reducing the ellipticity. In contrast, higher scan speeds led to increased eccentricities due to lower fluence and correspondingly smaller material modifications. These trends are illustrated in Fig.~\ref{fig:Fig4}.

\subsection{Waveguide width and modification dependence on laser fabrication parameters}

The repetition rate used in this work, 100~kHz, was selected based on the parameters reported by Imai \textit{et al.}~\cite{ref10}, which are comparable to the experimental conditions employed here. At this repetition rate, heat accumulation has not been observed in fused silica~\cite{ref3}, as discussed in the previous section. However, the formation of bright and dark dots was observed in waveguides fabricated at higher pulse energies and lower scanning speeds, corresponding to increased laser fluence. Although the pulse width could not be varied in the present study, Imai \textit{et al.} have shown that the density of these dots increases with longer pulse durations. \par

\begin{figure*}[t]
\centering
\includegraphics[width=\textwidth]{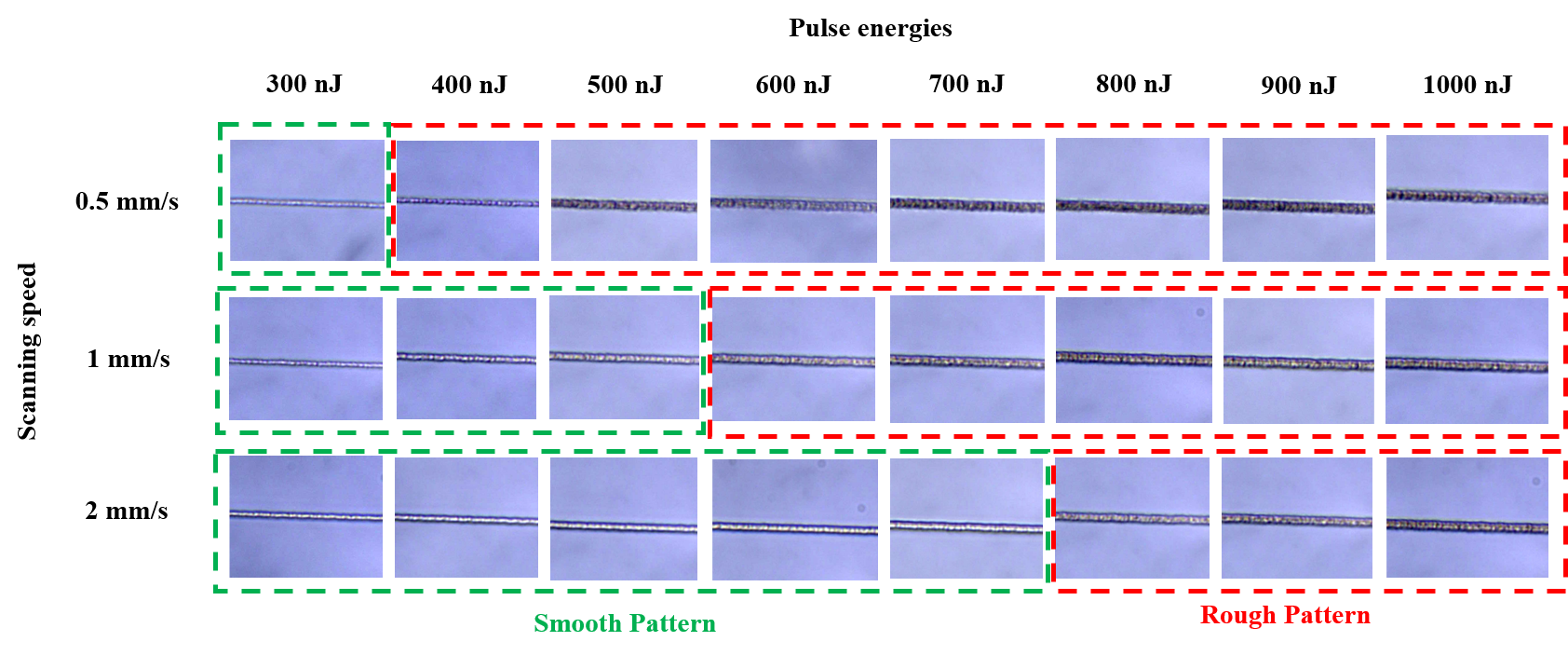}
\caption{Longitudinal optical microscope images of optical waveguides fabricated with different pulse energies and scanning speeds}
\label{fig:Fig5}
\end{figure*}

\begin{figure}[t]
\centering
\includegraphics[width=\columnwidth]{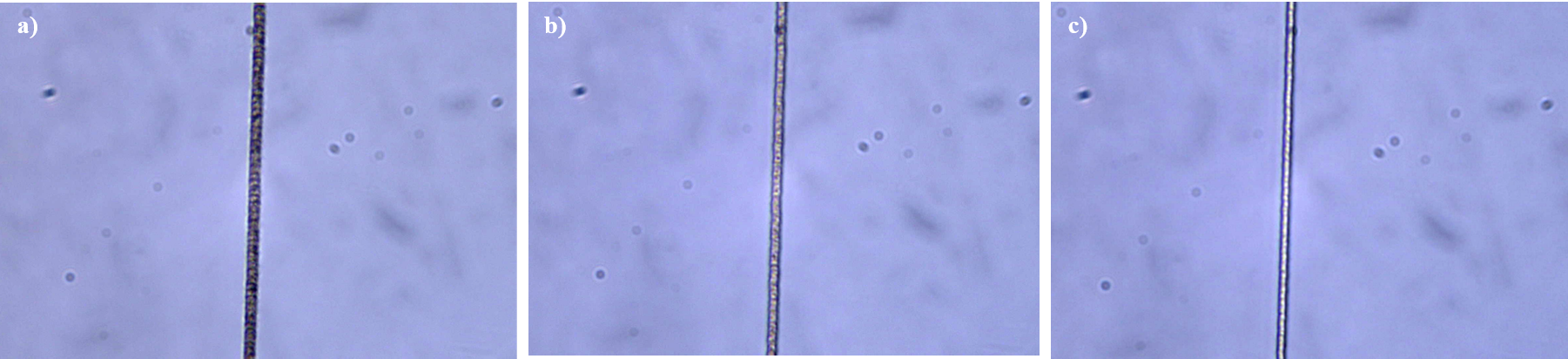}
\caption{Optical microscope images of waveguides fabricated at 600 nJ with scanning speeds of a) 0.5 mm/s, b) 1 mm/s, and c) 2 mm/s}
\label{fig:Fig6}
\end{figure}

It was observed that for a scan speed of 0.5~mm/s and pulse energies below 400~nJ, the fabricated waveguides were smooth and transparent. Similarly, for a scan speed of 1~mm/s, smooth waveguides were obtained for pulse energies below 600~nJ, while for a scan speed of 2~mm/s, the waveguides remained smooth for pulse energies below 800~nJ. This parametric investigation enabled the classification of the fabricated waveguides into rough-patterned and smooth-patterned categories for different combinations of fabrication parameters, as shown in Fig.~\ref{fig:Fig5}. In addition, Imai \textit{et al.} reported the formation of nano-wrinkles or nano-pores in the cross-sections of rough-patterned waveguides, which lead to strong optical scattering and consequently high propagation losses. The dependence of rough-pattern formation on scanning speed for a given pulse energy is illustrated in Fig.~\ref{fig:Fig6}. The waveguides were observed to be opaque, semi-transparent, and transparent for scanning speeds of 0.5~mm/s, 1~mm/s, and 2~mm/s, respectively, indicating that a scan speed of 2~mm/s is optimal for the fabrication of low-loss waveguides. \par

\begin{figure}[t]
\centering
\includegraphics[width=\columnwidth]{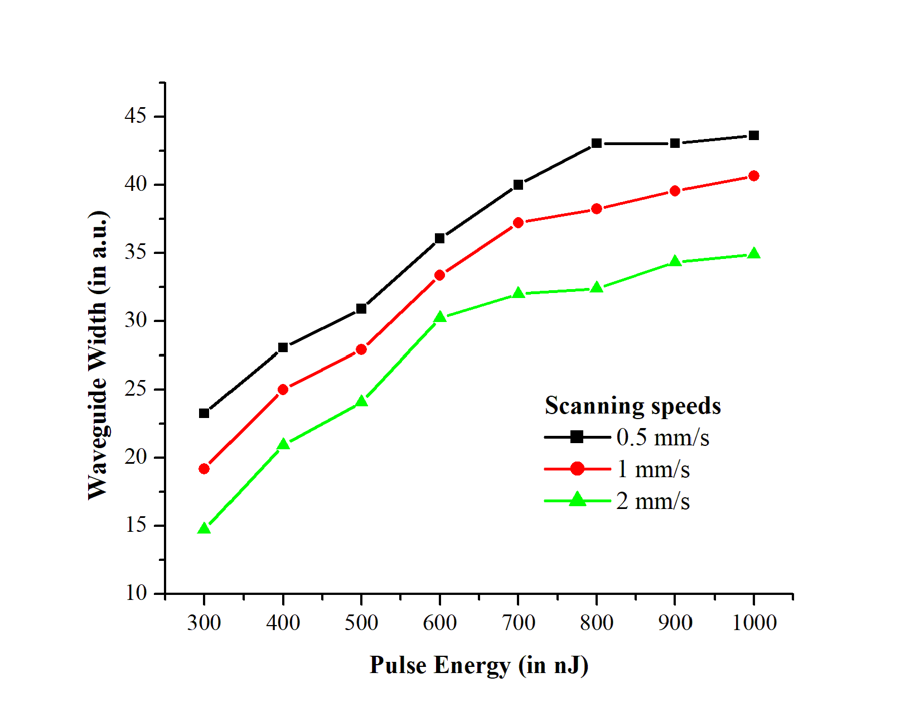}
\caption{Waveguide width as a function of pulse energies for different scanning speeds}
\label{fig:Fig7}
\end{figure}

The waveguide width was characterized as a function of pulse energy and scanning speed, as shown in Fig.~\ref{fig:Fig7}. A linear increase in waveguide width with increasing pulse energy was observed. Furthermore, for higher scan speeds, the waveguide widths were found to be smaller than those obtained at lower scan speeds. This behavior can be attributed to the fact that higher laser fluence, corresponding to slower scan speeds, induces larger material modifications, resulting in increased waveguide widths.

\subsection{Propagation loss characterization}

\begin{figure}[t]
\centering
\includegraphics[width=\columnwidth]{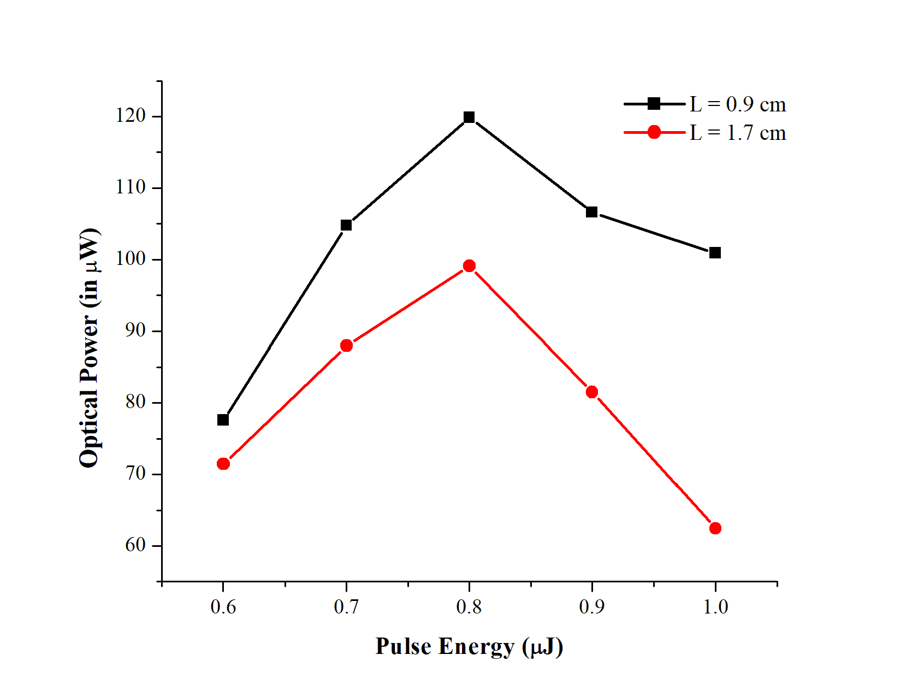}
\caption{Waveguide width as a function of pulse energies for different scanning speeds}
\label{fig:Fig8}
\end{figure}

\begin{figure}[t]
\centering
\includegraphics[width=\columnwidth]{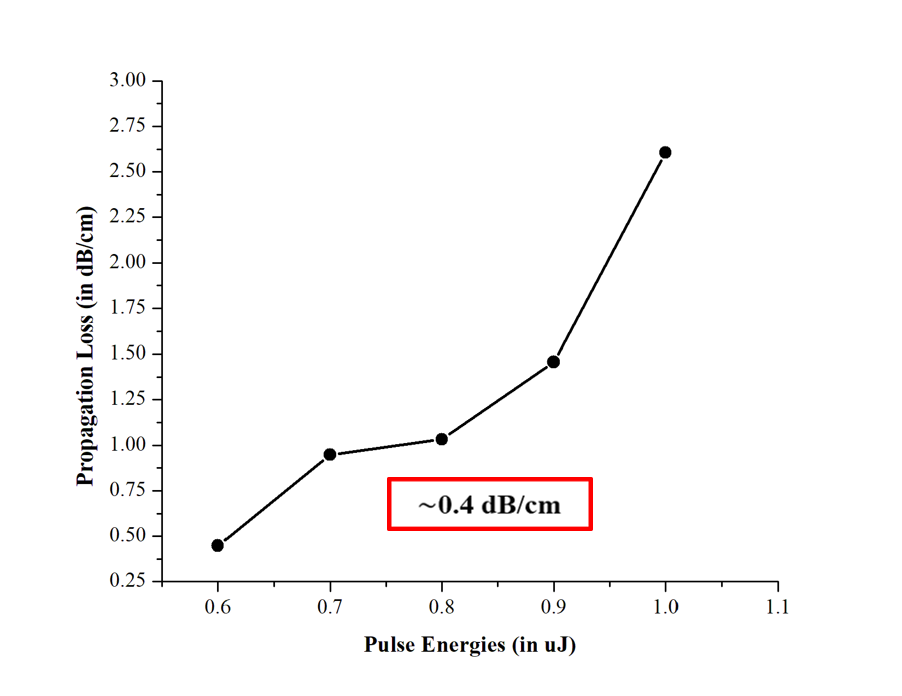}
\caption{Propagation loss as a function of pulse energies}
\label{fig:Fig9}
\end{figure}

To evaluate the quality of the fabricated waveguides, we measured the propagation loss of the smooth-patterned waveguides by varying the pulse energies from 600~nJ to 1000~nJ at an optimized scanning speed of 2~mm/s. As discussed in the previous section, we used the cut-back method, measuring the output power of two identical waveguides with different lengths as shown in Fig.~\ref{fig:Fig8}. For all the smooth-patterned waveguides, the propagation loss was $\leq$1~dB/cm, with the lowest being 0.4~dB/cm at a pulse energy of 600~nJ, as shown in Fig.~\ref{fig:Fig9}. \par
We did not perform propagation loss characterization for pulse energies below 600~nJ, as the goal was to use these waveguides for biosensing applications, which require a significant amount of light to be coupled into the waveguide. Although waveguides fabricated with pulse energies below 600~nJ would have lower propagation loss, they would also have a smaller numerical aperture, resulting in reduced light coupling, which is not ideal for our applications. Therefore the waveguides were fabricated with a combination of 600 nJ pulse energy and 2 mm/s scanning speed for our recent work \cite{ref22}.\par
According to previous reports, a propagation loss of less than 1~dB/cm is considered good for femtosecond laser micromachined waveguides. Imai \textit{et al.}\cite{ref10} report a propagation loss of $<$1~dB/cm in fused silica using the fundamental beam ($\lambda = 1028$~nm) with a pulse width of $<$400~fs and pulse energies between 710~nJ and 840~nJ. The parameters used in their study are very similar to those used here, which explains the similarity in propagation loss.\par
Amorim \textit{et al.}\cite{ref23} studied the loss mechanisms in femtosecond laser micromachined waveguides, identifying the influence of Rayleigh and Mie scattering on propagation loss, with reported coefficients of 0.5~dB/cm and 0.2--0.65~dB/cm, respectively. In another work, Florea et al. \cite{ref24} report propagation losses between 1.3 and 2.5~dB/cm in straight waveguides at 632.8~nm.

\subsection{Modal analysis of elliptical core waveguides}

It is well known that linearly polarized (LP) modes in circular-core waveguides or fibers are not exact solutions of Maxwell’s equations. The exact solutions are hybrid modes, typically represented as a superposition of $\mathrm{HE}_{mn}$, $\mathrm{EH}_{mn}$, $\mathrm{TE}_{0m}$, and $\mathrm{TM}_{0m}$ modes. However, in step-index fibers under the weakly guiding approximation ($n_1 \approx n_2$), the LP modes can be considered approximate solutions. These modes are useful for understanding and analyzing field distributions and transmission characteristics. In a circular-core waveguide or fiber, LP modes are represented by solutions of Bessel functions in the core region and modified Bessel functions in the cladding \cite{ref25}. According to Xia \emph{et al.} \cite{ref26}, mode degeneracy and instability in mode orientation are observed in circular-core waveguides and fibers.\par
In contrast, elliptical-core waveguides or fibers exhibit stable modal intensity distributions. The exact eigenvalue equations for an elliptical-core waveguide with arbitrary core and cladding refractive indices are given in \cite{ref27}. The modes in an elliptical-core waveguide or fiber are expressed as combinations of even and odd Mathieu functions in the core, and even and odd modified Mathieu functions in the cladding. Due to their stable modal characteristics, these waveguides and fibers find applications in optical communications and optical sensing.\par
By applying appropriate boundary conditions at the core--cladding interface, the modal characteristic equations can be derived. From these equations, the normalized cut-off values are determined after obtaining the general expressions for the cut-off frequencies and propagation constants. Notably, the cut-off frequencies of several higher-order modes depend strongly on the ellipticity of the core. Varshney \emph{et al.} \cite{ref28} reported that the normalized frequency parameter, or V-number, of an elliptical-core waveguide depends on the ratio of the semi-major axis to the semi-minor axis, commonly referred to as the aspect ratio. The V-number for an elliptical-core waveguide is given by
\begin{equation}
V_1 = \frac{2\pi a'}{\lambda} \sqrt{n_1^2 - n_2^2}, \quad n_1 > n_2,
\end{equation}
and
\begin{equation}
V = \frac{b'}{a'} V_1 
= \frac{\lambda b'}{2\pi a'^2} \frac{1}{\sqrt{n_1^2 - n_2^2}},
\end{equation}
where $a'$ and $b'$ are the semi-major and semi-minor axes of the elliptical core, respectively. According to Xia \emph{et al.} \cite{ref25}, when the aspect ratio $a/b > 2.475$, the $\mathrm{LP}_{11}$ even mode becomes the first higher-order mode, the $\mathrm{LP}_{21}$ even mode becomes the second, and the $\mathrm{LP}_{11}$ odd mode becomes the third higher-order mode in an elliptical-core waveguide. \par

Linearly polarized (LP) modes are particularly relevant in the context of femtosecond laser micromachined waveguides, as the induced refractive index change is typically on the order of the third decimal place. This small index contrast leads to the weakly guiding approximation, under which a scalar treatment of the wave equation is valid and LP modes provide a good approximation of the actual modal solutions. To gain further insight into the modal behavior, we also investigated these waveguides using numerical simulations.\par
The cross-section of an elliptical-core waveguide was modeled using COMSOL Multiphysics, and the supported modes were calculated using the ARPACK mode solver. The refractive indices of the core and cladding were assumed to be 1.457 and 1.45, respectively. The aspect ratio of the elliptical core was varied to study the number and nature of modes supported by the waveguide. The simulation results were consistent with the theoretical predictions reported by Xia \emph{et al.} \par
\begin{figure}[t]
\centering
\includegraphics[width=\columnwidth]{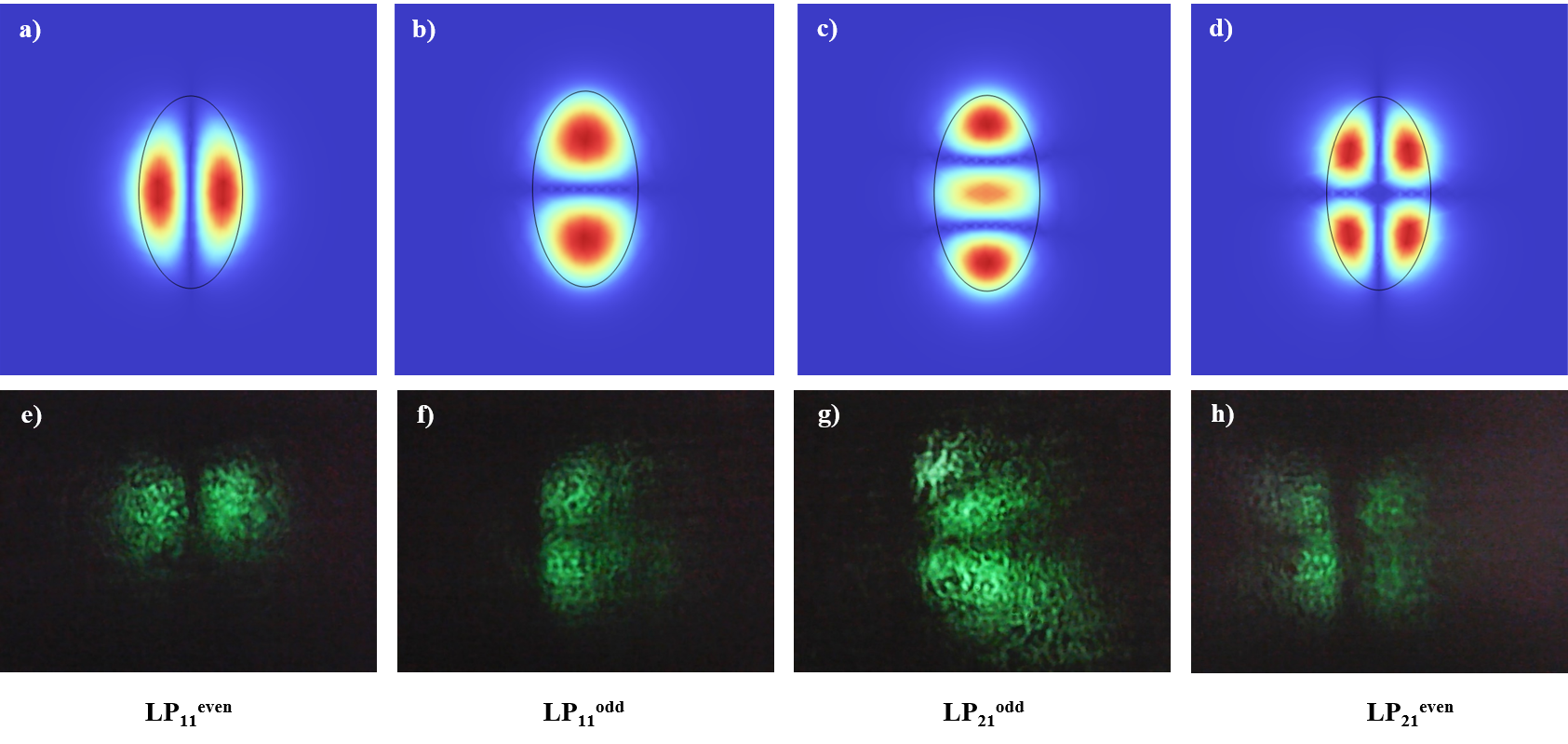}
\caption{a) – d) Computationally generated elliptic-core linearly polarized (LP) modes illustrating LP11even, LP21even, LP11odd, and LP21odd modes respectively; e) – h) Experimentally observed modes in fs-laser micromachined elliptical core waveguides.}
\label{fig:Fig10}
\end{figure}

To estimate the ratio of the major axis ($a$) to the minor axis ($b$), ImageJ software was used. Precise mode profiling was not considered necessary, as only the ratio $b/a$ was required for determining the core eccentricity. It was therefore assumed that any errors in the estimation of the major and minor axes through visual inspection would largely cancel out. We observed that the $\mathrm{LP}_{21}$ even mode indeed became the second higher-order mode following the $\mathrm{LP}_{11}$ even mode when the aspect ratio exceeded 2.475. This observation further confirmed the accuracy of the visual inspection method used to estimate the eccentricity. As shown in Fig.~\ref{fig:Fig10}, the experimentally observed mode profiles of the femtosecond laser micromachined elliptical-core waveguides closely resemble the expected profiles obtained from simulations. However, the micromachined waveguide modes were found to be slightly non-uniform in shape and orientation. This asymmetry can be attributed to the complex and non-uniform refractive index modification across the cross-section of the micromachined waveguides. A detailed investigation of the experimental mode shapes and their dependence on laser fabrication parameters requires further study and is left for future work. The primary objective of this study was to fabricate low-loss elliptical-core waveguides and explore their potential applications in bimodal optical biosensing.

\section{Conclusion}

In conclusion, elliptical-core waveguides were successfully fabricated using femtosecond laser micromachining in the first modification regime, characterized by a smooth refractive index change. The fabricated waveguides were analyzed for eccentricity and width variations as a function of pulse energy and scanning speed using bright-field microscopy. Propagation loss characterization was performed using the cut-back method, revealing that an optimized combination of 600~nJ pulse energy and a scanning speed of 2~mm/s resulted in a minimum propagation loss of 0.4~dB/cm. This low propagation loss demonstrates the high quality of the fabricated waveguides and confirms their suitability for practical applications. Consequently, this optimized set of laser fabrication parameters was consistently employed in all subsequent experiments, as it supported both the fundamental $\mathrm{LP}_{01}$ mode and the $\mathrm{LP}_{11}$ even mode, which are essential for bimodal interferometric operation. Numerical modal analysis of the elliptical-core waveguides was carried out using COMSOL Multiphysics, and the simulation results showed good agreement with established theoretical models reported in the literature. Experimentally, mode profiles obtained by launching a 532~nm wavelength source into the fabricated waveguides closely matched the simulated results, thereby validating the numerical analysis. Overall, this study demonstrates that precise control of femtosecond laser fabrication parameters enables the realization of low-loss elliptical-core waveguides, which are promising candidates for optofluidic biosensing applications.

\begin{acknowledgments}
The authors would like to thank Bhagawan Sri Sathya Sai Baba, the Founder Chancellor of SSSIHL, for being a constant source of inspiration, and for all the research facilities provided. They would also like to thank the Sri Sathya Sai Central Trust and DST-FIST (sanction no. SR/FST/PSI-172/2012) for the FEMTO FabULLAS facility at Central Research Instruments Facility (CRIF), SSSIHL. The authors also acknowledge support from the support from DST-Technology Development Program (DST-TDP-BDTD/15/2018).

\end{acknowledgments}

\section*{Data Availability Statement}
The data that support the findings of this study are available from the corresponding authors upon reasonable request.

\section*{References}
\bibliographystyle{apsrev4-1}
\bibliography{Bibliography}

\end{document}